\documentclass[twocolumn]{aastex631}

\def\Msun{M_\odot}
\def\MJ{M_\mathrm{J}}
\def\Mp{M_\mathrm{p}}
\def\Macc{\dot{M}_\mathrm{acc}}
\def\RJ{R_\mathrm{J}}
\def\Rp{R_\mathrm{p}}

\def\Lacc{L_\mathrm{acc}}
\def\LHa{L_\mathrm{H\alpha}}
\def\FHa{F_\mathrm{H\alpha}}
\def\AHa{A_\mathrm{H\alpha}}
\def\Av{A_\mathrm{V}}
\def\MaccAv{\overline{\dot{M}_\mathrm{acc}}}

\begin{document}

\title{MagAO-X and HST high-contrast imaging of the AS209 disk at H$\alpha$
}

\author[0000-0001-7255-3251]{Gabriele Cugno}
\affiliation{Department of Astronomy, University of Michigan, Ann Arbor, MI 48109, USA}

\author[0000-0003-2969-6040]{Yifan Zhou}
\affiliation{Department of Astronomy, The University of Texas at Austin, 2515 Speedway, Stop C1400, Austin, TX 78712, USA}

\author[0000-0003-4507-1710]{Thanawuth Thanathibodee}
\affiliation{Institute for Astrophysical Research, Department of Astronomy, Boston University, Boston, MA 02215, USA}

\author{Per Calissendorff}
\affiliation{Department of Astronomy, University of Michigan, Ann Arbor, MI 48109, USA}

\author{Michael R. Meyer}
\affiliation{Department of Astronomy, University of Michigan, Ann Arbor, MI 48109, USA}

\author{Suzan Edwards}\affiliation{Five College Astronomy Department, Smith College, Northampton, MA 01063, USA}

\author[0000-0001-7258-770X]{Jaehan Bae}
\affiliation{Department of Astronomy, University of Florida, Gainesville, FL 32611, USA}

\author[0000-0002-7695-7605]{Myriam Benisty}
\affiliation{Laboratoire Lagrange, Université Côte d’Azur, CNRS, Observatoire de la Côte d’Azur, 06304 Nice, France} 
\affiliation{Univ. Grenoble Alpes, CNRS, IPAG, 38000 Grenoble, France}

\author{Edwin Bergin}
\affiliation{Department of Astronomy, University of Michigan, Ann Arbor, MI 48109, USA}

\author[0000-0003-1863-4960]{Matthew De Furio}
\affiliation{Department of Astronomy, University of Michigan, Ann Arbor, MI 48109, USA}

\author[0000-0003-4689-2684]{Stefano Facchini}\affiliation{Universit\`a degli Studi di Milano, via Giovanni Celoria 16, 20133 Milano, Italy}

\author{Jared R. Males}
\affiliation{Steward Observatory, University of Arizona, Tucson, AZ 85179, USA}

\author{Laird M. Close}
\affiliation{Steward Observatory, University of Arizona, Tucson, AZ 85179, USA}

\author{Richard D. Teague}
\affiliation{Department of Earth, Atmospheric, and Planetary Sciences, Massachusetts Institute of Technology, Cambridge, MA 02139, USA}

\author{Olivier Guyon}
\affiliation{Steward Observatory, University of Arizona, Tucson, AZ 85179, USA}
\affiliation{James C. Wyant College of Optical Sciences, University of Arizona, 1630 E University Blvd, Tucson, AZ
85719, USA}
\affiliation{National Astronomical Observatory of Japan, Subaru Telescope, National Institutes of
Natural Sciences, Hilo, HI 96720, USA}
\affiliation{Astrobiology Center, National Institutes of Natural Sciences, 2-21-1 Osawa, Mitaka, Tokyo, JAPAN}

\author[0000-0001-5130-9153]{Sebastiaan Y. Haffert}\altaffiliation{NASA Hubble Fellow}
\affiliation{Steward Observatory, University of Arizona, Tucson, AZ 85179, USA}

\author{Alexander D. Hedglen}
\affiliation{James C. Wyant College of Optical Sciences, University of Arizona, 1630 E University Blvd, Tucson, AZ
85719, USA}

\author{Maggie Kautz}
\affiliation{James C. Wyant College of Optical Sciences, University of Arizona, 1630 E University Blvd, Tucson, AZ
85719, USA}

\author[0000-0001-8446-3026]{Andr\'es Izquierdo}\affiliation{European Southern Observatory, Karl-Schwarzschild-Str. 2, 85748 Garching bei München, Germany}
\affiliation{Leiden Observatory, Leiden University, P.O. Box 9513, 2300 RA Leiden, The Netherlands}

\author[0000-0003-1905-9443]{Joseph D. Long}
\affiliation{Steward Observatory, University of Arizona, Tucson, AZ 85179, USA}

\author{Jennifer Lumbres}
\affiliation{James C. Wyant College of Optical Sciences, University of Arizona, 1630 E University Blvd, Tucson, AZ
85719, USA}

\author{Avalon L. McLeod} 
\affiliation{James C. Wyant College of Optical Sciences, University of Arizona, 1630 E University Blvd, Tucson, AZ
85719, USA}

\author[0000-0003-3904-7378]{Logan A. Pearce}
\affiliation{Steward Observatory, University of Arizona, Tucson, AZ 85179, USA}

\author{Lauren Schatz}
\affiliation{Air Force Research Laboratory, Directed Energy Directorate, Space
Electro-Optics Division, Starfire Optical Range, Kirtland Air Force Base, NM, 87117,USA}

\author{Kyle Van Gorkom}
\affiliation{Steward Observatory, University of Arizona, Tucson, AZ 85179, USA}








\begin{abstract}

The detection of emission lines associated with accretion processes is a direct method for studying how and where gas giant planets form, how young planets interact with their natal protoplanetary disk and how volatile delivery to their atmosphere takes place. H$\alpha$ ($\lambda=0.656~\mu$m) is expected to be the strongest accretion line observable from the ground with adaptive optics systems, and is therefore the target of specific high-contrast imaging campaigns. 
We present MagAO-X and HST data obtained to search for H$\alpha$ emission from the previously detected protoplanet candidate orbiting AS209, identified through ALMA observations. 
No signal was detected at the location of the candidate, and we provide limits on its accretion. Our data would have detected an H$\alpha$ emission with $\FHa>2.5\pm0.3 \times10^{-16}$~erg~s$^{-1}$~cm$^{-2}$, a factor 6.5 lower than the HST flux measured for PDS70~b \citep{Zhou2021}. The flux limit indicates that if the protoplanet is currently accreting it is likely that local extinction from circumstellar and circumplanetary material strongly attenuates its emission at optical wavelengths.
In addition, the data reveal the first image of the jet north of the star as expected from previous detections of forbidden lines. Finally, this work demonstrates that current ground-based observations with extreme adaptive optics systems can be more sensitive than space-based observations, paving the way to the hunt for small planets in reflected light with extremely large telescopes.

\end{abstract}



\section{Introduction} \label{sec:intro}

In recent years observations by the Atacama Large Millimeter and Submillimeter Array (ALMA) and high contrast imagers including SPHERE and GPI revealed that circumstellar disks are highly structured, with gaps, rings and spirals being among the most important observable features \citep[e.g.,][]{Andrews2018, Avenhaus2018, Tschudi2021, Bae2022_rev, Benisty2022}. While other explanations have been proposed \citep[e.g., snowlines or turbulence, ][]{Zhang2015, Flock2015}, it is widely accepted that at least some of these structures are the result of the interaction with forming planets. 
This hypothesis is supported by the direct detection of two confirmed protoplanets in the cavity of the transition disk PDS70 \citep{Keppler2018, Haffert2019}. However, despite significant observational efforts, no other confirmed planet in the cavities of other disks has been detected, neither in the infrared \citep{AsensioTorres2021, Cugno2023} nor in emission lines associated to accretion \citep{Cugno2019, Zurlo2020, Xie2020, Huelamo2022, Follette2023}. Two candidates have been directly detected through H$\alpha$ emission \citep[LkCa15~b and AB~Aur~b,][]{Sallum2015, Currie2022}; the first has never been redetected, and near-infrared observations suggest it could be a reprocessed scattered light feature \citep{Currie2019}, while the second one shows H$\alpha$ emission consistent with scattered light from the disk \citep{Zhou2022} and requires further investigation.

An alternative method for inferring the existence of protoplanets and studying their formation is to examine the effect they have on the disk structure \citep{Pinte2022}. Indeed, young gas giant planets leave observational traces on the velocity field of the surrounding gas due to their gravitational force, manifesting as kinks \citep{Pinte2018, Pinte2020}, doppler flips \citep{Casassus2019}, deviations from Keplerian velocity \citep{Teague2018, Teague2021}, or molecular emission line broadening along the planet orbit \citep{Dong2019, Izquierdo2022}.

AS209 (alternative name V\,1121\,Oph) is a young ($\approx1-2$~Myr, \citealt{Andrews2009}) K5 ($V=11.3$ mag) star with a mass of $1.2~M_\odot$ \citep{Oberg2021, Teague2021}. \cite{Fernandez1995} measured the H$\alpha$ flux of the star to be $8.4\times10^{-12}$~erg~s$^{-1}$~cm$^{-2}$, which translates in a mass accretion rate of $\Macc=10^{-8.2}~\Msun$~yr$^{-1}$ when using the line luminosity vs. accretion luminosity relationship from \cite{Fang2009}. This value is in strong agreement with the mass accretion rate of $\Macc=10^{-8.3}~\Msun$~yr$^{-1}$ obtained by \cite{Fang2018} using multiple emission line luminosities other than H$\alpha$. It is surrounded by a well-studied protoplanetary disk ($i=35\pm0.1^\circ$, \citealt{Huang2018}) that has been the subject of several major surveys such as DSHARP \citep{Andrews2018, Huang2018, Guzman2018}, DARTTS-S \citep{Avenhaus2018}, and MAPS \citep{Oberg2021}. As part of the MAPS program, \cite{Bae2022} identified a candidate circumplanetary disk (CPD) detected in $^{13}$CO within a gap at $\sim200$~au previously identified in $^{12}$CO and scattered light observations \citep{Guzman2018, Avenhaus2018}. Subsequent kinematic analysis of the molecular line data revealed disk winds emerging from the gap around the candidate, showing a complex interplay between forming planets and disk winds \citep{Galloway2023, Izquierdo2023}. A wind from the central star was also inferred through high resolution spectroscopy of forbidden lines \citep{Fang2018, Banzatti2019}. 

\begin{figure*}[t!]
\centering
\includegraphics[width=0.99\hsize]{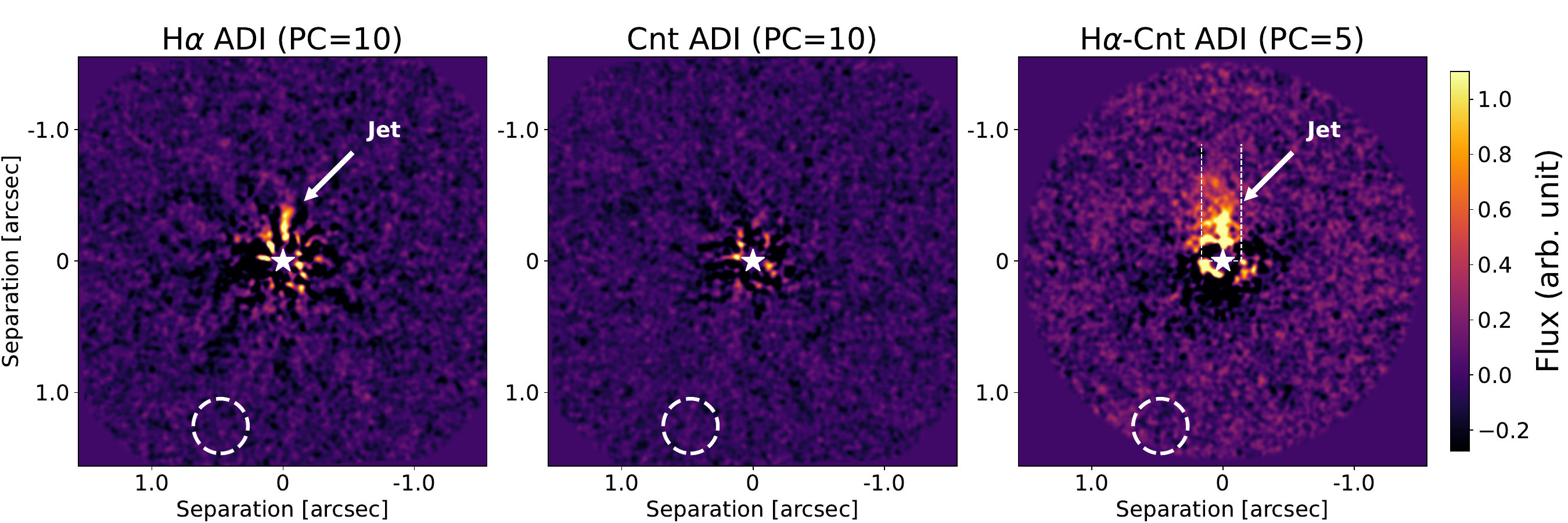}
\caption{MagAO-X AS209 residuals in the H$\alpha$ {(\it left panel)}, in the continuum {(\it middle panel)}, and in the continuum-subtracted H$\alpha$ frames {(\it right panel)}. North points to the top and East to the left. The white star indicates the position of the central star. The position of the CPD candidate is marked with a white dashed circle. The jet can be identified as the extended emission North of the star in both the left and the right panels. The rectangular mask applied for the jet brightness measurement is shown in the right panel.
\label{fig:AS209c}}
\vspace{0.5cm}
\end{figure*}

In this paper we search for H$\alpha$ emission from the the CPD candidate orbiting AS209 with the Magellan Adaptive Optics eXtreme (MagAO-X)  instrument \citep{Males2022} and with HST/WFC3/UVIS. MagAO-X is a new high-contrast imaging instrument that pushes extreme adaptive optics to visible wavelengths \citep{males2018magaox, close2018magaox, Males2022}, including H$\alpha$ \citep{close2020separation}. MagAO-X is a successor to MagAO \citep{Close2014_spie} and has an increased number of controllable modes (1600 versus 300), AO loop speed (2 kHz versus 1 kHz) and sensitivity (twice the throughput at H$\alpha$). The data reveal a stellar jet at short separations, while no localized H$\alpha$ emission at the location of the CPD candidate or anywhere else within the disk is detected.
This Paper is structured as follows: in Sect.~\ref{sec:observations} we present the MagAO-X and HST observations and in Sect.~\ref{sec:reduction} we detail our data reductions. The main results are reported in Sect.~\ref{sec:results} and discussed in Sect.~\ref{sec:discussion}. The conclusions of our work can be found in Sect.~\ref{sec:conclusions}.

\section{Observations}\label{sec:observations}

\subsection{MagAO-X}\label{sec:MagAO_observations}

AS209 was observed with the MagAO-X instrument at the Las Campanas Observatory on April 17 2022 and April 20 2022. The observing conditions were very different between the two nights. On April 17, data were taken under very stable conditions, with seeing $<0\farcs5$ for most of the night, airmass$<1.2$ and photometric sky. 
Conversely, due to extremely unstable atmospheric conditions on April 20, the second set of data were not usable. As a result, this manuscript focuses solely on the data obtained on April 17. 

AS209 was observed in H$\alpha$ dual-band imaging mode, which involves splitting the light into the H$\alpha$ ($\lambda_c=0.656~\mu$m, $\Delta\lambda=9$~nm) and nearby continuum (Cnt, $\lambda_c=0.668~\mu$m, $\Delta\lambda=9$~nm) filters after it has undergone the same optical path through the instrument. This results in the very similar diffraction and speckle pattern in both filters, allowing the continuum image to be used to remove the stellar contribution from the H$\alpha$ image without affecting any potential line flux emitted from a protoplanet. Typical protoplanets do not contribute in the continuum image at $\sim0.65~\mu$m due to their relatively low temperatures ($\sim1000-2000$~K) compared to stars. For more details on this observational mode and the reasoning behind it we refer to \cite{Close2014} and \cite{Cugno2019}.

We obtained a total of 209 frames in each filter with a Detector Integration Time (DIT) of 60~s per exposure. Thanks to the high sky rotation rate, we achieved a total field rotation of $\sim106^\circ$, ensuring a high throughput for the angular differential imaging \citep[ADI;][]{Marois2006} post-processing algorithm.

\begin{figure}[t!]
\centering
\includegraphics[width=0.99\hsize]{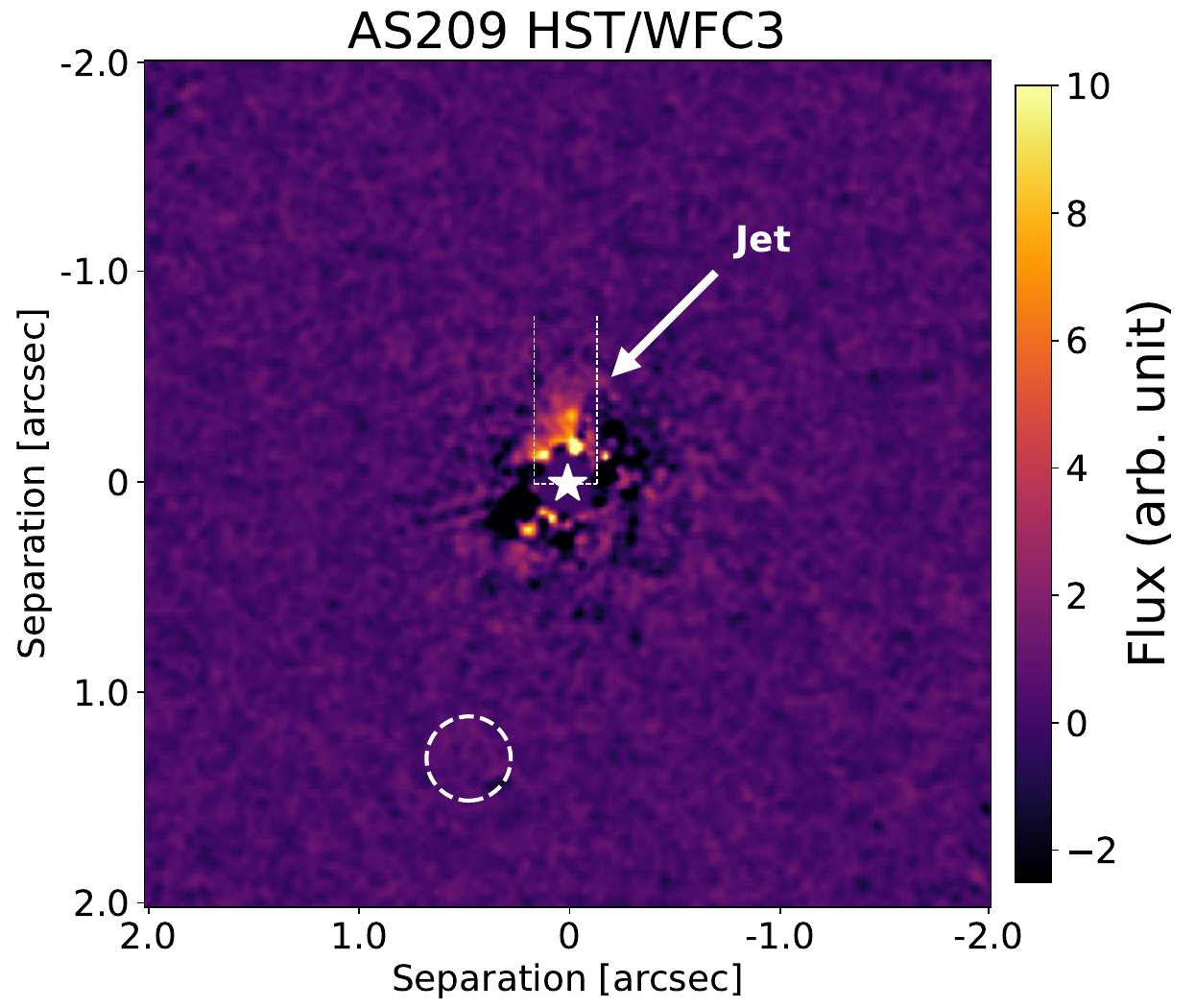}
\caption{HST AS209 residuals in the F656N filter (H$\alpha$). North points to the top and East to the left. The white star indicates the position of the central star, while the position of the CPD candidate is marked with a white dashed circle. The jet North of the star is re-detected in agreement with MagAO-X data. The rectangular mask applied for the jet brightness measurement is shown.
\label{fig:AS209c_HST}}
\vspace{0.5cm}
\end{figure}

\subsection{HST/WFC3/UVIS}\label{sec:HST_observations}
AS209 was observed with the HST/WFC3/UVIS instrument on 2023-05-03 for three orbits in the F656N (H$\alpha$ narrowband, $\lambda_c=6561.5$~\AA, $\Delta\lambda=17.9$~\AA) filter using the UVIS2/C512C subarray (field of view $20\farcs5\times20\farcs5$). To improve the spatial sampling, a four-points dithering pattern was used, in which the telescope moves by 0.5 pixels between exposures. In total we obtained 108 frames of 13~s each, for a total time on target amounting to 23.4 min. In order to perform ADI, the roll angle in Orbit 2 differs by 35$^\circ$ from those in Orbits 1 and 3.

\section{Data reduction} \label{sec:reduction}

\subsection{MagAO-X}\label{sec:MagAO_reduction}

The MagAO-X data were reduced using the high-contrast imaging pipeline {\tt PynPoint} \citep{Amara2012, Stolker2019}. After dark and flat calibration, frames are flipped along the x-axis to correct for a reflection within the instrument and bad pixels are corrected by 4-sigma clipping. Images are then aligned to each other using cross-correlation and then centered by fitting a 2D Gaussian function to the mean image. Continuum and H$\alpha$ frames were found to display an azimuthal offset of about $0.3^\circ$ and we corrected this by rotating the continuum images. This process resulted in squared images of $3\farcs7$ in size (MagAO-X pixel scale is $0\farcs0059$). 

We applied a frame selection based on the peak intensity of the PSF in the H$\alpha$ filter, measured in an aperture of radius $0\farcs0135$ (2.3 pixels) in order not to be biased by the position of the star on the pixel grid. Since the instrument response to a point source is the same for every source, when the stellar PSF has a lower peak due to low Strehl, the same applies to faint protoplanet signals, making them more difficult to detect. We removed images with peak fluxes that deviated by $>2\sigma$ from the maximum value of the entire dataset. This step removed 45 frames out of 209 from the dataset, most of them temporally located towards the end of the night, when weather conditions started to deteriorate.

Two different PSF subtraction techniques were applied. First, we removed the stellar PSF from individual filters (both H$\alpha$ and Continuum) using ADI based on principal component analysis \citep[PCA,][]{Amara2012}. We found that at least 10 principal components have to be subtracted in order to remove the bright stellar noise. Second, the PSF was removed using a combination of dual-filter differential imaging \citep{Close2014, Cugno2019} followed by ADI.  
The continuum frames are first spatially downscaled to match the PSF size (which is $\propto \lambda$) at the H$\alpha$ wavelength\footnote{Information on the MagAO-X H$\alpha$ filters can be found at \url{https://magao-x.org/docs/handbook/observers/filters.html}} and their flux is upscaled by a factor 2.1 to match the total flux in the corresponding H$\alpha$ image within an aperture of $r=1\farcs0$. This is necessary to compensate for the substantial line emission contribution due to stellar accretion in the H$\alpha$ filter. We found that the continuum subtraction accurately cancels most of the stellar noise, and only a few components have to be removed to reveal faint sources.
After the final residuals were median-combined, we applied a Gaussian filter of the size of the PSF (FWHM=0\farcs034) to reduce pixel-to-pixel variation and highlight protoplanetary candidates.

\subsection{HST}\label{sec:HST_reduction}
We initiated the data reduction process using the \texttt{flc} file obtained from the MAST archive. To begin with, we visually inspected the data in order to identify any hot pixels and cosmic rays present. These pixels were then replaced by employing linear interpolations based on the neighboring pixels. 
Subsequently, we reconstructed Nyquist sampled images by interlacing sets of four dithered images in Fourier space (for details on the image reconstruction we refer the reader to \citealt{Lauer1999,Zhou2021}). This step resulted in 27 Nyquist sampled images with pixelscale $0\farcs02$. 

For the primary subtraction, we again employed {\tt PynPoint}. To align the images, we registered them according to the centroids of the PSFs, which were determined through two-dimensional Gaussian fitting. The centroid-aligned image cube was then fed into the PCA algorithm, where images from one roll position are used to model and subtract the central PSF in the other roll position. For HST data, only a few components were removed. Finally, after derotating and median-combining the frames we applied again the highpass filter (FWHM$=0\farcs04$).

\section{Results}\label{sec:results}

The residual images  (shown in Fig.~\ref{fig:AS209c} and Fig.~\ref{fig:AS209c_HST}) do not reveal a signal at the location of the CPD candidate \citep[$1\farcs4$,][]{Bae2022}, and in Sect.~\ref{sec:AS209b} we calculate detection limits at the candidate location and discuss the non-detection. In addition, the residuals show extended emission North of the star coming from a jet both in the MagAO-X and HST data, presented in Sect.~\ref{sec:jet}.

\begin{figure*}[t!]
\includegraphics[width=\hsize]{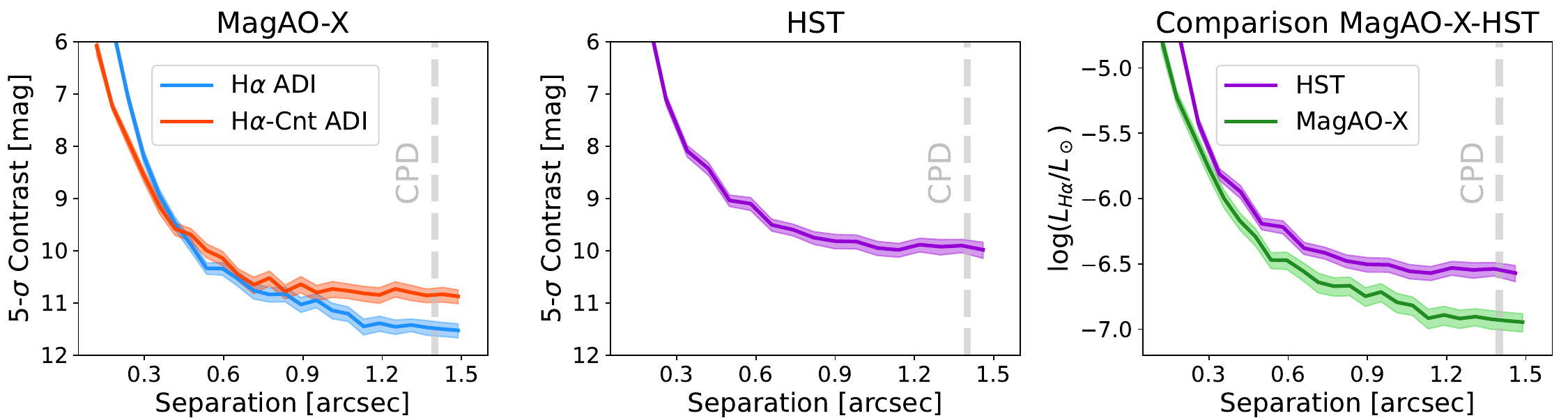}
\caption{H$\alpha$ $5\sigma$ contrast limits and H$\alpha$ luminosity limits of AS209 as a function of separation. The grey dashed lines represent the separation of the CPD candidate \citep{Bae2022}. Shaded regions represent the systematic uncertainty due to the speckle and residuals noise. {\it Left:} contrast curves obtained with MagAO-X. The curve obtained with the H$\alpha$ differential technique is shown with an orange line, while the curve obtained with ADI PSF-subtraction without the continuum removal is shown with a blue line. Subtracting the scaled continuum improves the contrast performance by up to 1~mag, especially at short separations ($\lesssim0\farcs3$). {\it Middle:} contrast curves obtained with HST. Due to different stellar fluxes in the filters used, the contrasts from the left and central panels are not directly comparable. {\it Right:} H$\alpha$ luminosity upper limits as a function of separation for the two instruments: the curves are directly comparable, indicating that our MagAO-X data are more sensitive than our HST data at every separation in the image. 
\label{fig:limits}}
\end{figure*}

\subsection{The CPD candidate}\label{sec:AS209b}
The protoplanet candidate surrounded by a circumplanetary disk detected in $^{13}$CO by \cite{Bae2022} was not detected neither with MagAO-X nor with HST in H$\alpha$. Its expected position is shown in the residual images of Fig.~\ref{fig:AS209c} and Fig.~\ref{fig:AS209c_HST}. 

To quantify the relevance of the non detection, we estimated detection limits on the presence of H$\alpha$ emitting sources. These were obtained with the {\tt applefy} tool presented in \cite{Bonse2023} based on the metric proposed by \cite{Mawet2014}. The detection threshold was fixed to a false positive fraction (FPF) of $2.87\times10^{-7}$, equivalent to 5$\sigma$ for large separations and Gaussian noise. Artificial protoplanetary signals were inserted every  $0\farcs06$ at four different PAs. The planet signal and the noise were measured in an aperture of $r=3.2$~pix and 2.0~pix for MagAO-X and HST data (half FWHM), respectively. For every separation the noise for 360 different aperture placements is estimated and we report the median over all results with the standard deviation representing the systematic uncertainty on the contrast measurement (see \citealt{Bonse2023} for more details).  Because the presence of the bright jet could bias the contrast curves, especially at short separations, we rotated the residuals in the opposite direction before combining them and estimating the noise \citep{Pairet2019}. This approach maintains the stellar and speckle noise, but avoids the presence of physical signals in the images used to sample the noise. 

The obtained $5\sigma$ contrast curves from MagAO-X data are shown in the left panel of Fig.~\ref{fig:limits}, in orange for the H$\alpha-$Cnt  and in blue for the H$\alpha$ stack with PCA-ADI PSF-subtraction. For the injection of planets we used the PSF from the continuum filter, as about 25\% of the data in the H$\alpha$ filter is saturated in the inner few pixels. In addition, stellar line emission may suffer from variability on different timescales and therefore the H$\alpha$ flux is more difficult to calibrate. We note that the different filter throughputs might introduce a bias in the flux estimate of a few percent, much smaller than the contrast uncertainties. 
At small separations, H$\alpha$ differential imaging outperforms ADI, providing contrasts $\sim1$ mag deeper. At larger separations, simple ADI reaches higher contrasts than H$\alpha$ differential imaging. We attribute this behavior to the higher detector and readout noise introduced in the data by the continuum subtraction in the outer regions of the images, where the stellar emission does not dominate.

The contrast limits of the left panel of Fig.~\ref{fig:limits} show that at the separation of the CPD candidate we reach a 5$\sigma$ contrast of $11.5\pm0.1$~mag. For the absolute calibration, we adopted $F^*_\mathrm{Cnt}=1.01~(\pm0.1)\times10^{-11}$~erg~s$^{-1}$~cm$^{-2}$ for the stellar flux in the continuum filter of the MagAO-X data \citep{Henden2015}, where the 10\% uncertainty was conservatively chosen to include potential variability (the nominal measurement uncertainty is 4\%). 
Combining this value with the contrast limit measured for the CPD candidate, we obtained a $5\sigma$ H$\alpha$ flux limit for the protoplanet of $F^\mathrm{H\alpha}_\mathrm{pl} \lesssim 2.5\pm0.3 \times10^{-16}$~erg~s$^{-1}$~cm$^{-2}$, which once corrected for the stellar distance ($d=121\pm0.4$~pc, \citealt{Gaia2022}) corresponds to a luminosity of $L_\mathrm{H\alpha}^\mathrm{obs} \lesssim 1.15\pm0.15 \times 10^{-7}~L_\odot$. The overall luminosity upper limit as a function of separation from the MagAO-X data is shown in green in the right panel of Fig.~\ref{fig:limits}. At each separation, we adopted the higher contrast between the two curves reported in the left panel.

The contrast limits measured by HST are shown in the central panel of Fig.~\ref{fig:limits}. At the separation of the CPD candidate we reach a contrast of $9.9\pm0.1$~mag. The contrast measured with respect to the stellar flux in the only HST filter available is not directly comparable to the one obtained above for the MagAO-X data for two reasons: (i) the HST filter is centred on the H$\alpha$ line, while for a more reliable flux calibration the MagAO-X contrast limits were measured with respect to stellar flux obtained in the continuum filter, and (ii) the HST filter has a width of $18$~\AA, 5$\times$ smaller than the MagAO-X filters ($\sim90$~\AA). From the HST images, we were able to directly measure the stellar flux in the F656N filter: $3.22\pm0.05\times10^{-13}$~erg~s$^{-1}$~cm$^{-2}$~\AA$^{-1}$.Applying the contrast at the separation of the CPD candidate, we obtain a flux limit from the HST data of $F^\mathrm{H\alpha}_\mathrm{pl} \lesssim 5.7\pm0.6 \times10^{-16}$~erg~s$^{-1}$~cm$^{-2}$, and once corrected for stellar distance a luminosity upper limit of $L_\mathrm{H\alpha}^\mathrm{obs} \lesssim 2.6\pm0.3 \times 10^{-7}~L_\odot$. These two quantities are directly comparable to the MagAO-X data, which provided lower limits and are able to exclude planets with dimmer luminosities. Hence, for the interpretation of the non-detection we will only focus on the upper limits provided by MagAO-X. The comparison of sensitivities achieved by the two observatories is further discussed in Sect.~\ref{sec:comparison}.

\begin{figure}[ht!]
\centering
\includegraphics[width=0.99\hsize]{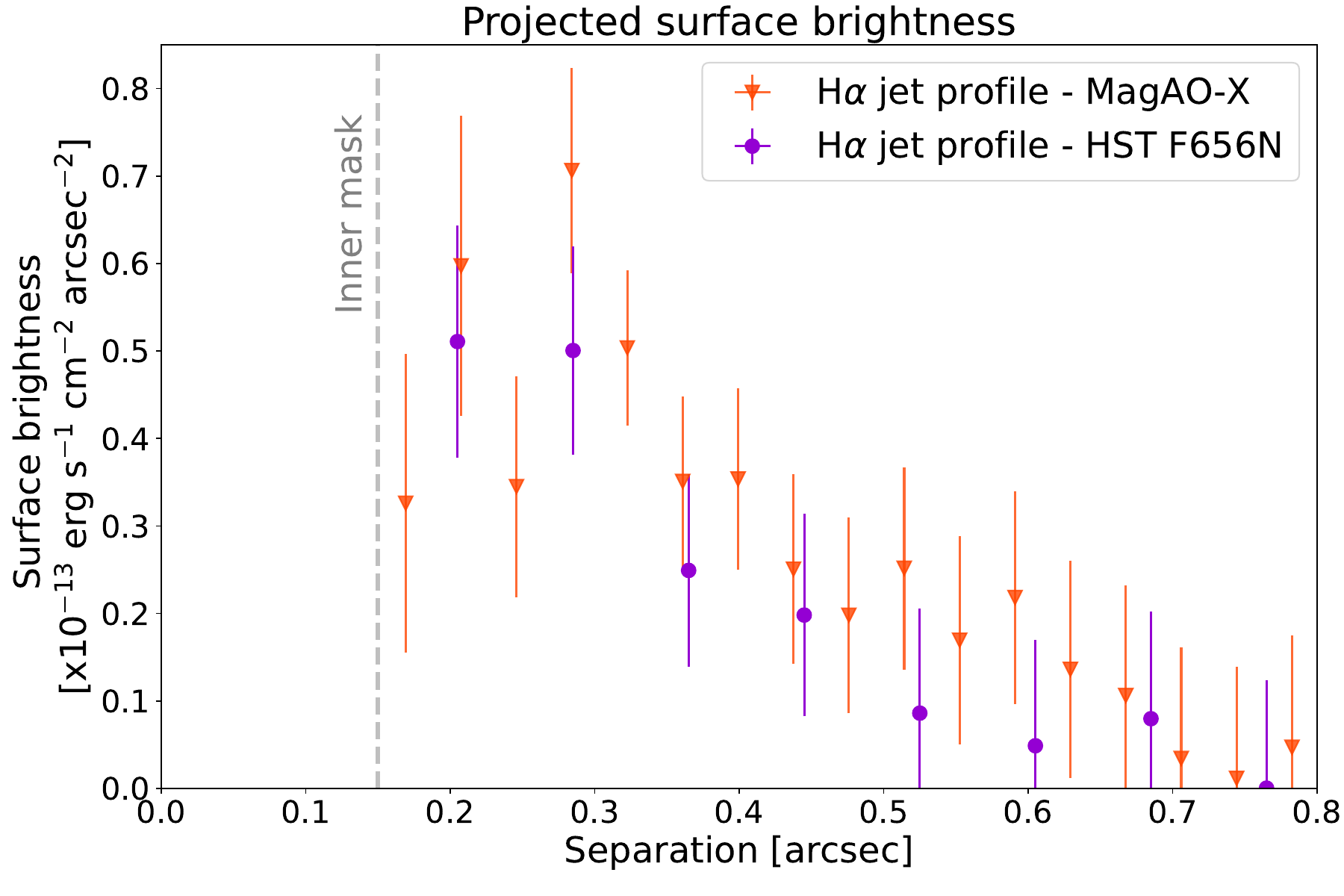}
\caption{Surface brightness profile of the jet from AS209 as measured from MagAO-X (orange) and HST (violet) data. Errorbars represent $1\sigma$ uncertainties. The central mask with radius $0\farcs15$ is shown with the dashed grey line.
\label{fig:profile}}
\vspace{0.5cm}
\end{figure}

\subsection{Jet}\label{sec:jet}
Figure~\ref{fig:AS209c} and Figure~\ref{fig:AS209c_HST} reveal a jet north of the star imaged for the first time in AS209.
Its extension in the 2D projected image goes up to a separation of $0\farcs65$ in the MagAO-X data. The jet in the HST data is detected only up to $0\farcs5$, thus the values provided next only refer to MagAO-X data. Assuming the jet is perpendicular to the disk plane, its physical extension is estimated to be $\sim169$~au. Its width in the residuals is $\sim0\farcs3$, equivalent to $\sim36$~au at AS209 distance. This value is in line with FWHM of jets from other TTauri stars \citep{Ray2007}. No receding jet is detected, possibly due to the obscuring effect of the circumstellar disk.

To estimate the jet profile, we applied a rectangular mask ( width~$=0\farcs3$ for both instruments) to focus only on the image region with the jet signal (see dashed lines in the right panel of Fig.~\ref{fig:AS209c} and in Fig.~\ref{fig:AS209c_HST}). To avoid the more aggressive subtraction induced by PCA, we used residuals produced with classical ADI: given the large field rotation in the MagAO-X data and the roll angle observing strategy used to obtain the HST data, we expect subtraction of the extended jet signal to be minimal (throughput $\sim100\%$) at large enough separations. However, some level of self-subtraction is expected in the innermost region of the images. In addition, those same regions suffer from strong speckles residuals from the PSF subtraction. Hence, we masked the inner $0\farcs15$ in each dataset. The flux from the the jet was measured in annuli in steps of 1 FWHM ($\sim6.5$~pix and $\sim4$~pix for MagAO-X and HST data respectively), and normalized to the measurement area, obtaining the surface brightness profile. To calibrate the flux, we estimated the PSF counts within an aperture of $2\farcs0$ and we compare it to the stellar flux as described in Sect.~\ref{sec:AS209b} to find a count-to-flux conversion factor for each instrument. The obtained conversion factors were applied to the brightness profiles, and the calibrated profiles are reported in Fig.~\ref{fig:profile}. The profiles estimated from the two instruments are in agreement. For the errorbars, we considered two types of uncertainties. The first one, corresponding to the noise induced by stellar residuals, was estimated by considering the residuals outside the jet mask at the same separations and assigning an uncertainty to the measurement corresponding to the fraction of the noise counts over the jet counts (normalized by area). The second one corresponds to photon noise. At each separation we considered the largest source of uncertainty (residuals $\lesssim0\farcs35$, photon noise $\gtrsim0\farcs35$). The profiles obtained with the two instruments largely agree with each other.

The jet appears rather smooth and does not show significant knots as it was observed for RY Tau \citep{Garufi2019, Uyama2022} and HD163296 \citep{Ellerbroek2014, Xie2021}. However, it seems some intensity variation is present between $0\farcs15$ and $0\farcs3$ in the MagAO-X data, even though with the current data it is difficult to assess if the bumps in the profile are physical or due to the post-processing.

\section{Discussion}
\label{sec:discussion}

\subsection{The protoplanet mass}
\label{sec:IR}
AS209 has been observed with VLT/NaCo in the $L'$-band as part of the NaCo-ISPY survey \citep{Launhardt2020}. The data, presented in \cite{Cugno2023}, exclude the presence of a companion brighter than 17.2 mag, which translates in an absolute magnitude of 11.8 mag at $L'$. 
Even though \cite{Cugno2023} showed the risk of using evolutionary models to transform IR brightness measurements into mass estimates, we use the detection limits to obtain a rough estimate of what near-IR high-contrast imaging can exclude, assuming a clear view of the planet photosphere and no contribution of the accretion luminosity at $L'$.
For hot-start scenarios like AMES-Cond, NaCo limits exclude objects with $\Mp>1.5~\MJ$, a very tight upper limit. However, for a colder scenario like the for the {\tt BEX-Warm} models presented in \cite{Marleau2019} the mass limit is $10.6~\MJ$ (assuming an age of $\tau=2$~Myr), providing a much more loose constraint. We note that a planet with such a high mass would have opened a much deeper and wider gap in the gas distribution of the disk than observed in $^{12}$CO \citep{Kanagawa2015}.

\cite{Bae2022} used an empirical planet mass $-$ gap width relation and calculated the mass of the planet embedded in the CPD candidate to be $0.42-4.2~\MJ$, depending on the disk viscosity, assuming it is responsible for carving out the gas gap at $\sim200$~au. We note however that more recent work suggested that even lower masses are possible from a dynamical point of view. In fact, \cite{Galloway2023} found disk winds emerging from the gap the CPD candidate is embedded in, and suggested that the kinematics could be dominated by the winds, not the planet. In this case, the planet mass inferred from the gap width could be over-estimated. These mass estimates agree with the VLT/NaCo detection limits.

\subsection{The non-detection from the candidate}

Different models can be used to transform measured H$\alpha$ luminosities to mass accretion rates. Here, we compare two sets of models: the emission model of shock-heated gas for planetary mass objects \citep{Aoyama2018, Aoyama2020}, and the magnetospheric accretion model applied to planetary objects presented in \cite{Thanathibodee2019}.
Based on their non-Local Thermal Equilibrium (non-LTE) models, \cite{Aoyama2021} determined the $L_\mathrm{H\alpha}-L_\mathrm{acc}$ relation for protoplanets. Applying the relation to our measurement from Sect.~\ref{sec:AS209b}, we obtain $\log(L_\mathrm{acc}/L_\odot)=-6.16\pm0.05$. Following \cite{Gullbring1998}, the mass accretion rate $\Macc$ can then be estimated with 
\begin{equation}
    \Macc = \left(1-\frac{\Rp}{R_{in}}\right)^{-1} \frac{\Lacc \Rp}{G \Mp},
    \label{eq:macc}
\end{equation}
where $\Rp$ is the planet radius, $R_{in}$ is the disk truncation radius, assumed to be $\approx5\Rp$, and $G$ is the gravitational constant. The mass of the planet has been assumed to be between $0.5~\MJ$ and $4~\MJ$ (see Sect.~\ref{sec:IR}) and radii were drawn uniformly between 1 and $3~\RJ$.

In order to consider magnetospheric accretion, we use the models from \cite{Thanathibodee2019} to generate a large grid of scenarios producing H$\alpha$ lines. For each calculation, we systematically varied the planets' and model's parameters encompassing a wide range of values whose boundaries are given either by the AS209 system possible properties or by extreme scenarios. Using masses between 0.5 and 4~$\MJ$, we adopted evolutionary models from \cite{Mordasini2017} to estimate radii for each mass consistent with 1-2~Myr ($\Rp$ in the $1.9-3.1~\RJ$ range). For a given set of mass and radius, we calculate a grid of 4050 models varying  the magnetospheric parameters. The mass accretion rate $\log(\Macc/\MJ$~yr$^{-1})$ was taken between $-9$ (weak H$\alpha$ produced) to $-5$ (typical stellar accretion rate), the maximum flow temperature from 7000~K to 8000~K (lower temperature don't produce H$\alpha$)\footnote{We note this parameter is very uncertain even for the more studied case of accreting stars \citep{Muzerolle2001}.}, the magnetospheric truncation radius from $1.5~\Rp$ (smaller values would bring the disk too close to the planet) to $6~\Rp$ (larger values require an unlikely strong magnetic field), flow width from $0.5~\Rp$ to $2.5~\Rp$ (similar to stellar scenario), and the inclination from 20 to 50 degree (consistent with the circumstellar disk's inclination). We then calculate the line luminosity from each model and select the model from which the luminosity is within 1$\sigma$ of the upper limits.

\begin{figure}[b!]
\includegraphics[width=\hsize]{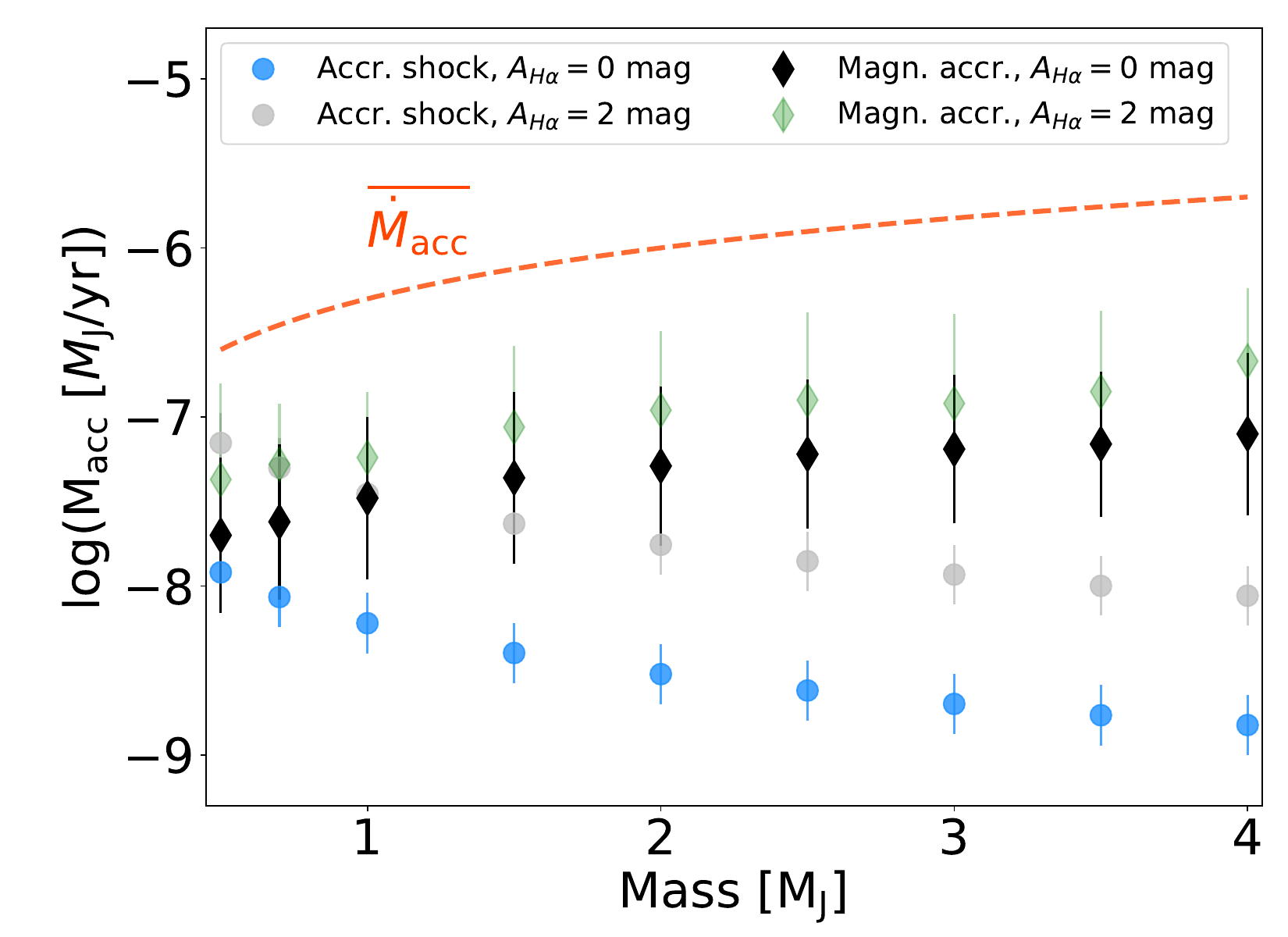}
\caption{Mass accretion rate upper limit of the protoplanet candidate as a function of the planet mass $\Mp$ under different scenarios. The orange dashed line represents the average mass accretion necessary to build a planet of a given mass in 2 Myr. The blue and grey circles represent the mass accretion rate estimated using the shock-heated accretion models from \cite{Aoyama2021} using the observed H$\alpha$ luminosity ($\AHa=0,2$ respectively). Black and green diamonds indicate the mass accretion rate when applying the magnetospheric accretion model for $\AHa=0$~mag and 2 mag respectively. 
\label{fig:mass_accretion_rate}}
\end{figure}

Initially, we did not consider extinction ($\AHa=0$~mag). Figure~\ref{fig:mass_accretion_rate} shows the mass accretion rate estimate as a function of the planet mass for the H$\alpha$ limit of the CPD candidate. The orange dashed line represents the average mass accretion rate $\MaccAv$ necessary to reach the planet mass on the $x$-axis in 2~Myr. As it is still unclear if and how accretion rates onto protoplanets evolve with time and on which timescales, we use $\MaccAv$ as a proxy for a constant mass accretion rate over the stellar lifetime (2 Myr). If $\Macc>\MaccAv$, then the planet is either undergoing a period of vigorous (above average) accretion (meaning it underwent lower accretion in the past) or the planet age is lower than 2~Myr. If $\Macc<\MaccAv$, then a large fraction of the planet mass was accreted in the past and currently the mass accretion rate is significantly lower than earlier phases. In Fig.~\ref{fig:mass_accretion_rate} the blue circles represent the distribution of possible $\Mp-\Macc$ using the shock-heated gas models, while the black diamonds report the results from the magnetospheric accretion models. 

The two models show different trends for increasing planet masses. The reason is that in the  magnetospheric accretion model, to the first order, the line flux scales with the flow density $\rho\propto\dot{M}/\sqrt{M_p}$ \citep[, Eq.~9]{Hartmann1994} and hence for fixed line fluxes $\rho$ does not vary and $\Macc\propto\sqrt{\Mp}$. Conversely, in the accretion shock models free fall velocity of the infalling material is higher for a more massive planet, increasing the H$\alpha$ flux. As a consequence, to produce the same H$\alpha$ emission lower $\Macc$ is required at higher masses. The comparison betwen $\MaccAv$ and the results for both sets of models suggests that the accretion rate is $10-100$ times lower than $\MaccAv$. Hence, assuming no extinction, it seems the planet is accreting at a relatively low rate, indicating that in the past the accretion rate was likely much more vigorous. 

However, circumstellar and circumplanetary material can strongly affect line emission at H$\alpha$ wavelengths, especially in the presence of dust. \cite{Avenhaus2018} detected scattered light from the disk gap in AS209 indicative of the presence of small grains. Moreover, the non-zero disk inclination ($i=35^\circ$, \citealt{Huang2018}) possibly increases its effect on the detectable line emission. 
Unfortunately, the extinction properties of the outer regions of a circumstellar disk have never been measured, not to mention that we expect it to vary between sources and within individual sources depending on the location with respect to substructures and disk geometry. 

From the non-detection of H$\beta$ and the re-detection of H$\alpha$, \cite{Hashimoto2020} constrained the extinction  towards PDS70\,b to be $\AHa\gtrsim2$~mag, while \cite{Uyama2021} inferred $A_V\approx0.9$ and $2.0$~mag for PDS70\,b and c respectively considering the non-detection of Pa$\beta$. Following these estimates, we adopt here $\AHa=2.0$~mag at the position of the CPD candidate, as it is located, similar to the PDS70 planets, in a deep gap, where the amount of small grains and pebble is expected to be relatively low (but not absent, see \citealt{Avenhaus2018}). 

Including the effect of the extinction from the circumstellar disk material in the gap shifts the results from both models vertically along the $y$-axis of Fig.~\ref{fig:mass_accretion_rate}. The new mass accretion rates are now closer to $\MaccAv$, but still $1-2$ orders of magnitude lower. If this is the case, we conclude again that the protoplanet candidate accreted material much more vigorously in the first phases of its life \citep{Lubow2012, Brittain2020}. The consequence of this early accretion scenario for direct imaging surveys of H$\alpha$ emission would be drastic: at younger ages, when planets would accrete at larger rates, stars are still embedded in their envelopes and the direct detection of line emission would likely be hindered by the much larger extinction in the optical. 

We note however, that the PDS70 planets (even though located in a different region of the disk) show signs of relatively strong accretion ($>10^{-7}~\MJ$~yr$^{-1}$, \citealt{Hashimoto2020, Zhou2021}) at much older ages \citep[$\tau\sim8$~Myr,][]{Wang2021}. If the protoplanet candidate continues to accrete material at least at a similar rate, an additional source of attenuation is needed to explain the non-detection, likely from circumplanetary material\footnote{We note that \cite{Marleau2022} showed that the accreting material only is not able to substantially increase the extinction along the line of sight.}. This scenario would be consistent with the interpretation of the $^{13}$CO detection from \cite{Bae2022}, in which a circumplanetary disk surrounds the forming planet. We note that, under the assumption of a planetary atmosphere with physical and chemical properties similar to those of more mature planets, \cite{Cugno2021} estimated the extinction from the circumplanetary environment necessary to suppress molecular features from VLT/SINFONI data of PDS70~b to be $\Av\approx10-15$ mag ($\AHa\approx8-12$~mag). For the same object, \cite{Wang2021} estimated the extinction from fitting the protoplanet's SED with atmospheric models to be up to $A_V\sim10$~mag. A similar value for the extinction towards the protoplanet candidate would allow the $\LHa$ upper limits to be consistent with $\MaccAv$. The much larger extinction from the CPD material is expected to strongly suppress the H$\alpha$ signal produced by accretion processes. If this is the case, accretion tracers at longer wavelengths (e.g., Br$\gamma$ at $2.16~\mu$m, Br$\alpha$ at $4.05~\mu$m) could help reducing the effect of the extinction and confirming the presence of an accreting CPD.

A comparison of the CPD candidate around AS209 with the PDS70 planets reveals significant environmental differences that could explain the non-detection of the former and the multiple detections of the latter. The more advanced developmental stage of the PDS70 system potentially indicates a different evolutionary phase, wherein the planets may be less embedded compared to much younger protoplanetary disks like AS209, as with larger masses and longer times they are more capable of clearing gaps \citep{Szulagyi2020, Sanchis2020}. Moreover, the presence of two planets inhabiting the same gap likely ensures an even higher level of gap clearing, thereby mitigating the effect of extinction on the emitted H$\alpha$ flux. 
In order to thoroughly characterize the evolution of forming planets, a greater number of confirmed protoplanet direct detections is necessary, but aforementioned dissimilarities suggest that distinct features are observable during the various phases of their formation.

\subsection{A jet in a planet-forming disk}
Images of microjets (extensions $\sim100$~au) in Class II sources are relatively rare and are mostly confined to forbidden lines \citep{Ray2007, Pascucci2022}. However, microjets from TTauri stars with high accretion rates are inferred to be common via spatially unresolved spectra of strong forbidden lines such as [O I] ($\lambda$6300) and [S II] ($\lambda$6731), where a `high velocity component' is attributed to collimated outflowing gas with centroids around $-100$~km~s$^{-1}$, with the red side of the line occulted by the disk \citep{Hartigan1995}. In contrast, detection of a microjet at H$\alpha$ in a TTauri star is usually prohibitive due to the high-contrast required to distinguish it from the broad emission arising from magnetospheric accretion onto the star. As an example, in AS209 the equivalent width of H$\alpha$ is 113 \text{\AA}  and the line wings extend to $\pm$ 300~km~s$^{-1}$ \citep{Alencar2000}. Instruments such as MagAO-X promise to be game changing for imaging microjets at H$\alpha$ in accreting Class II sources that, in tandem with forbidden lines, will enable physical parameters, such as density, temperature and mass ejection rates to be determined, as is currently done for resolved jets from Class I and 0 sources \citep[e.g.][]{Nisini2005}.

Of two published profiles of [O I] 6300 in AS209, one from KECK's HIRES spectrograph  \citep{Fang2018} and the other from Magellan's MIKE spectrograph \citep{Banzatti2019}, only the second shows a high velocity component associated with jet emission, while both show a low velocity component associated with a slow, extended disk wind. Whether this is due to jet variability or to different slit alignments is unclear. The accretion rate for AS209 has been estimated from calibrated relationships between emission lines and accretion luminosity established from Balmer Jump emission \citep{Alcala2017} to be $\log(\Macc/(M_\odot~\mathrm{yr}^{-1}))=-8.3$ \citep{Fang2018}, sufficiently high for jet emission to be expected \citep{Nisini2018}.

The receding jet was not detected, likely due to absorption from the protoplanetary disk material. Although receding jets are sometimes seen, \citep[for example the Ae/Be star HD163296, see ][]{Xie2021}, this is a relatively rare occurrence in TTauri stars. 

Future work with instruments like the Visible Integral-field Spectrograph eXtreme \citep[VIS-X,][]{Haffert2021} will be able to spectrally and spatially resolve the jet at H$\alpha$, allowing for the investigation of its velocity components. In addition, the calculation of  ratios with other emission lines in the jet will enable the estimation of the level of ionization and thus jet mass loss rate. Finally, an exciting prospect is the comparison between the stellar and the planetary mass accretion rates, once emission lines, possibly at longer wavelengths, will be detected.

\subsection{The future of ground-based high-contrast imaging in the optical}
\label{sec:comparison}
The detection limits reported in Fig.~\ref{fig:limits} show that in this case the data obtained with MagAO-X are more sensitive than those obtained from space with HST at every separation. Even though at least partially this is due to the very stable weather conditions during the observations and the longer amount of time spent on target (3.5 hr versus 23.4 min)\footnote{ Note that including overheads, the amount f time used for MagAO-X and HST observations is comparable, showcasing the efficiency of ground-based imaging.}, this work demonstrates the success of the recent efforts from the community to improve AO correction at short wavelengths. An additional point to consider is that AS209 is a relatively faint star ($V=11.3$~mag) and the AO-correction will be even better for brighter targets like nearby stars. This is an exciting premise for the future development of ground-based high-contrast imagers like ELT/EPICS \citep{Kasper2010} and GMagAO-X \citep{Males2022_elt} that in the next decade will search for small terrestrial planets in reflected light.

\section{Summary and Conclusions}
\label{sec:conclusions}
We have observed AS209 with the MagAO-X instrument on the 6.5m Magellan Clay telescope at Las Campanas Observatory in dual imaging mode and with the HST/WFC3/UVIS camera in the F656N filter. We detected a collimated jet in H$\alpha$, which can be spectrally resolved with future observations to study the different velocity components. In addition, no signal from the CPD candidate proposed in \cite{Bae2022} has been detected, suggesting that either the instantaneous mass accretion rate is very low, or that local extinction from the circumstellar and circumplanetary environments is strongly attenuating line emission in the optical.

The interpretation of the protoplanet candidate orbiting AS209 and its mass accretion rate is highly dependent on the flux absorption. 
Our current inability to accurately characterize the extinction from these elements prevents us from further constraining the accretion onto the protoplanet candidate. Future observations should try to detect other hydrogen recombination lines in the near-IR, where extinction plays a minor role. The comparison of line luminosities will further constrain the extinction from the line emitting region.

\begin{acknowledgments}
GC thanks the Swiss National Science Foundation for financial support under grant number P500PT\_206785. Y.Z. acknowledges support from the Heising-Simons Foundation 51 Pegasi b Fellowship. This project has received funding from the European Research Council (ERC) under the European Union’s Horizon 2020 research and innovation programme (PROTOPLANETS, grant agreement No. 101002188). 
SF is funded by the European Union under the European Union’s Horizon Europe Research \& Innovation Programme 101076613 (UNVEIL).
SYH was supported by NASA through the NASA Hubble Fellowship grant \#HST-HF2-51436.001-A awarded by the Space Telescope Science Institute, which is operated by the Association of Universities for Research in Astronomy, Incorporated, under NASA contract NAS5-26555.
LC and the MaxProtoPlanetS survey are supported by NASA eXoplanet Research Program (XRP) grant 80NSSC21K0397.
LAP acknowledges research support from the NSF Graduate Research Fellowship under Grant No. DGE-1746060.
The results reported herein include data gathered with the {\it Magellan Clay} 6.5-meter telescope located at Las Campanas Observatory, Chile. MagAO-X was developed under NSF MRI Award \#1625441.
The observations and data analysis works were supported by program HST-GO-17283. Some of the data presented in this paper were obtained from the Mikulski Archive for Space Telescopes (MAST) at the Space Telescope Science Institute. The specific observations analyzed can be accessed via \dataset[https://doi.org/10.17909/nf39-4h66]{https://doi.org/10.17909/nf39-4h66}. Supports for Program numbers HST-GO-17283 were provided by NASA through a grant from the Space Telescope Science Institute, which is operated by the Association of Universities for Research in Astronomy, Incorporated, under NASA contract NAS5-26555.

\end{acknowledgments}

%

\vspace{5mm}
\facilities{Magellan: Clay (MagAO-X); Hubble Space Telescope (WFC3)}


\software{PynPoint \citep{Stolker2019},  
          applefy \citep{Bonse2023}, 
          }






\bibliography{sample631}{}
\bibliographystyle{aasjournal}



\end{document}